\documentclass[aps,prd,twocolumn,floatfix,nofootinbib,superscriptaddress,tightenlines]{revtex4}
\usepackage[dvipsnames]{xcolor}
\usepackage{amsmath}
\usepackage{dcolumn}
\usepackage{lipsum}
\usepackage{amssymb}
\usepackage{soul}
\usepackage{url}
\usepackage{epsfig}
\usepackage{graphicx}
\usepackage{amsmath}
\usepackage{bm}
\usepackage{setspace}
\usepackage{appendix}
\usepackage{lscape}
\usepackage{amsthm}
\usepackage{bbold}
\usepackage{dcolumn}
\usepackage{epsfig}
\usepackage{graphics}
\usepackage{graphicx}
\usepackage{longtable}

\usepackage{bm}
\usepackage{xspace}
\usepackage{cancel}
\usepackage{float}
\usepackage{multirow}
\definecolor{darkgreen}{rgb}{0,0.5,0}
\definecolor{purple}{rgb}{0.5,0,0.5}
\definecolor{nblue}{rgb}{0.0,0.0,0.50}
\definecolor{scarlet}{rgb}{1.0,0.2,0}
\definecolor{darkmagenta}{rgb}{0.55, 0.0, 0.55}
\definecolor{darkolivegreen}{rgb}{0.33, 0.42, 0.18}
\definecolor{darkcandyapplered}{rgb}{0.64, 0.0, 0.0}

\usepackage[colorlinks=true, pdfstartview=FitV, linkcolor=purple, citecolor= purple, urlcolor=blue]{hyperref}

\newcommand{\be}{\begin{equation}}
\newcommand{\tu}{\textcolor{red}{u}}

\newcommand{\f}{\textcolor{blue}{f}}
\newcommand{\fu}{\textcolor{blue}{\bar{f_2}}}
\newcommand{\fd}{\textcolor{blue}{f_1}}

\newcommand{\Me}{\textcolor{blue}{V}}

\newcommand{\Jpsi}{\textcolor{blue}{J/\Psi}}

\newcommand{\td}{\textcolor{darkcandyapplered}{d}}
\newcommand{\tb}{\textcolor{blue}{b}}
\newcommand{\tc}{\textcolor{darkmagenta}{c}}
\newcommand{\ts}{\textcolor{darkgreen}{s}}
\newcommand{\ee}{\end{equation}}
\newcommand{\bea}{\begin{eqnarray}}
\newcommand{\eea}{\end{eqnarray}}
\newcommand{\beas}{\begin{eqnarray*}}
\newcommand{\eeas}{\end{eqnarray*}}
\newcommand{\nn}{\nonumber}

\newcommand{\MeV}{\text{MeV}} 
\newcommand{\GeV}{\text{GeV}} 

\newcommand{\tab}[1]{Table~\ref{#1}}

\begin{document}
\title{Electric, Magnetic and Quadrupole Form Factors and Charge Radii of Vector Mesons: \\ From Light to Heavy Sector in a Contact Interaction}
\author{R. J. Hern\'andez-Pinto}
\email{roger@uas.edu.mx}
\affiliation{Facultad de Ciencias F\'isico-Matem\'aticas, Universidad Aut\'onoma de Sinaloa, Ciudad Universitaria, Culiac\'an, Sinaloa 80000,
M\'exico}

\author{L. X. Guti\'errez-Guerrero}
\email{lxgutierrez@conahcyt.mx}
\affiliation{CONAHCyT-Mesoamerican Centre for Theoretical Physics,
Universidad Aut\'onoma de Chiapas, Carretera Zapata Km.~4, Real
del Bosque (Ter\'an), Tuxtla Guti\'errez, Chiapas 29040, M\'exico}

\author{M. A. Bedolla}
\email{marco.bedolla@unach.mx}
\affiliation{Facultad de Ciencias en Física y Matemáticas, Universidad Aut\'onoma de Chiapas, 
Carretera Emiliano Zapata Km.~8, Rancho San Francisco, Ciudad Universitaria Ter\'an, Tuxtla Guti\'errez, Chiapas 29040, M\'exico}

\author{A. Bashir}
\email{adnan.bashir@umich.mx 
\\ adnan.bashir@dci.uhu.es}

\affiliation{Instituto de F\'isica y Matem\'aticas, Universidad
Michoacana de San Nicol\'as de Hidalgo, Edificio C-3, Ciudad
Universitaria, Morelia, Michoac\'an 58040, M\'exico}
\affiliation{Department of Integrated Sciences and Center for Advanced Studies in Physics, Mathematics and Computation, University of Huelva, E-21071 Huelva, Spain}


\begin{abstract}
 We present a detailed survey of electric, magnetic and quadrupole form factors of light and heavy spin-1 vector mesons. It complements our analogous analysis of the electromagnetic form factors of pseudoscalar and scalar mesons reported earlier. Our formalism is based upon the Schwinger-Dyson equations treatment of a vector $\times$ vector contact interaction and the Bethe-Salpeter equation description of relativistic two-body bound states. We compute the form factors, associated moments and charge radii, comparing these quantities to earlier theoretical studies and experimental results if and when possible. We also investigate the quark-mass dependence of the charge radii and find the anticipated hierarchy such that it decreases with increasing dressed quark masses. In addition, our analysis shows that the magnetic moment is independent of the mass of the light and heavy mesons. Our results agree with most measurements reported earlier, finding a negative quadrupole moment, implying the charge distribution is oblate.

\end{abstract}

\maketitle

\section{Introduction}
Quantum chromodynamics (QCD), \cite{Wilczek:2012sb,Wilczek:1999be}, is a quantum field theory that describes sub-nuclear strong interactions in terms of the elementary degrees of freedom, namely, quarks and gluons within the celebrated standard model of Salam, Weinberg and Glashow. 
The infrared dynamics of QCD is responsible for nearly 98\% of the visible mass, dubbed as the emergent mass, in the universe. These interactions account for the formation of all strongly interacting bound states such as protons, neutrons, pions, $\rho$, etc. However, note that the ground-state spin-1 vector (V) mesons like $\rho$ are dissimilar to the pseudoscalar (PS) mesons such as pions which are Nambu-Goldstone bosons associated with dynamical chiral symmetry breaking and are consquently very light. The mass of the $\rho$ particle is about two-thirds that of a proton but almost six times heavier than a pion. In the chiral limit, where the Higgs field contribution to the light quarks is zero, a pion will be strictly massless. Therefore, we can conclude that the pion mass predominantly owes itself to the interference of emergent mass and the Higgs mechanism. On the other hand, the $\rho$ meson mass primarily arises from the emergent mass contribution. For heavier quarks and heavier mesons, the current quark masses and thus the Higgs mechanism dominates.



Thomas Jefferson National Accelerator Facility (JLab) measures the electromagnetic form factors (EFFs) of pions and kaons. Both the JLab's current and future upgrade as well as  the planned experiments at the Electron-Ion Collider (EIC) have the unprecedented potential to measure the EFFs of these PS mesons till much larger values of the probing photon momentum squared, i.e., $Q^2$. 
These efforts will contribute towards improving our understanding of the complex internal structure and charge distribution of these PS bound states, both in the non-perturbative and perturbative regimes,~\cite{Accardi:2023chb,Aguilar:2019teb}. However, for the V mesons, the situation is different. 
Brief lifetime of these mesons (approximately $4.5\times 10^{-24}$ sec for $\rho(\tu\bar{\td})$, 
$7.2\times 10^{-21}$ sec for $J/{\rm \Psi}(\tc\bar{\tc})$ and
$1.21\times 10^{-20}$ sec for $\Upsilon(\tb\bar{\tb})$) makes it challenging to extract their experimental EFFs, resulting in them naturally receiving less attention compared to PS mesons. On the other hand, V mesons can provide additional insights into the bound state that cannot be acquired from zero-spin mesons, such as the current distribution and the deviation from spherical symmetry through computing the quadrupole moment. It can also serve as a benchmark to study the
the deuteron, the lightest spin-1 nucleus, which has a unique structure that offers insights into complex nuclear interactions.
Again JLab and EIC are widely believed to be ideal facilities for studying the spin-1 deuteron,~\cite{Accardi:2012qut}. 
Moreover, in the quark-diquark picture
of the nucleon and its excited states,
axial vector (AV) diquarks play a significant role. These systems are the partner diquarks of
the $\rho$-meson and a better understanding of the $\rho$ leads to an improved knowledge of the AV diquark and its role in the study of baryons,~\cite{Barabanov:2020jvn}.

Within a coupled formalism based upon an infinite set of non-linear integral Schwinger-Dyson equations (SDEs)
and the Bethe-Salpeter equations (BSEs), electric, magnetic and quadrupole form factors of the light V mesons, $\rho$ and $K^*$, were computed in Refs.~\cite{Hawes:1998bz,Bhagwat:2006pu,Xu:2019ilh} with slightly varying degree of modeling sophistication and assumptions. These were also calculated through other
theoretical approaches such as  lattice QCD,~\cite{Owen:2015gva,QCDSF:2008tjq,Shultz:2015pfa}, holographic models,~\cite{Grigoryan:2007vg,Gurjar:2024wpq}, as well as Nambu--Jona-Lasinio (NJL) model~\cite{Zhang:2024nxl}. The
EFFs of heavy-light PS and V mesons within the SDE-BSE framework were recently reported in Ref.~\cite{Xu:2024fun}. 
The $\rho$ EFFs in the time-like region were studied in~\cite{deMelo:2016lwr}.
EFFs of heavy flavored V mesons $D_s, D_s^*, J/\Psi$ and $B_s, B_s^*, \Upsilon$ were computed in relativistic independent quark model in Ref.~\cite{Priyadarsini:2016tiu}.



In recent years, a symmetry preserving momentum independent vector$\times$vector contact interaction (CI) model has been employed within the SDE-BSE based formalism. Proposed initially in the Ref.~\cite{Gutierrez-Guerrero:2010waf}, it captures most essential features of infrared QCD. Since
its inception, it has been used to study a wide range of mesons, including PS, scalar (S), V, AV mesons and the corresponding diquark elastic and transition form factors 
in the light quark sector,~\cite{Roberts:2010rn,Roberts:2011wy,Chen:2012txa,Wang:2022mrh,Xing:2022jtt}. It was later extended to the study of static properties and form factors in the heavy quarks sector~\cite{Bedolla:2016yxq,Raya:2017ggu,Gutierrez-Guerrero:2019uwa} and also to the case of baryons~\cite{Wilson:2011aa,Segovia:2013uga,Segovia:2014aza,Raya:2018ioy,Raya:2021pyr}. More recently, the CI has also been employed to compute the gravitational form factors of the ground state PS mesons~\cite{Sultan:2024hep} and the pion Boer-Mulders function~\cite{Cheng:2024gyv}. 
In this article, we provide a comprehensive analysis of the  electric, magnetic and quadrupole form factors of all light and heavy
spin-1 V mesons in continuation of our recent work on the broad mass spectrum of ground state PS mesons
and their EFFs,~\cite{Pinto:2022tic,Hernandez-Pinto:2023yin}.

We have organized this article as follows: in Sec.~\ref{ingredients}, we present all the necessary tools and ingredients of the CI model required for the subsequent study of the V mesons.
It implements a symmetry-preserving regularization and the SDE formulation of the CI,
following Refs.~\cite{GutierrezGuerrero:2010md,Roberts:2011wy, Hernandez-Pinto:2023yin}.
Bethe-Salpeter Amplitude (BSA) for the V mesons is discussed and the masses computed. The quark-photon vertex is analyzed at length.
In Sec.~\ref{sec:eff} the generalities regarding the 
computation of the EFFs through the triangle diagram are detailed, including a discussion on the charge radii and analytic representation of the results obtained. In Sec.~\ref{dis}, we present our numerical results for the three electric, magnetic and quadrupole form factors along with the corresponding radii, supplemented with the detailed discussion on the numerical results obtained. 
Sec.~\ref{Conclusions} provides a summary and perspective about our results.
  Finally, some of the details of our elaborate calculation are given in the appendices \ref{ingredients} and \ref{App:EM}.

\section{The Contact Interaction} 
\label{ingredients}

Although most details of the CI can already be found in 
earlier references and literature, yet for the sake of completeness, and for detailing a particular mass-dependent choice of parameters for all ground state V mesons of a wide range of mass spectrum, we recall its most relevant features in this section, including the generation of dressed mass for the quarks through the gap equation and the V mesons via the bound state BSE.

\subsection{The gap equation: dressed quark masses}

\begin{figure}[hbt]
   \vspace{-3.7cm}
   \centering
    \includegraphics[scale=0.35,angle=0]{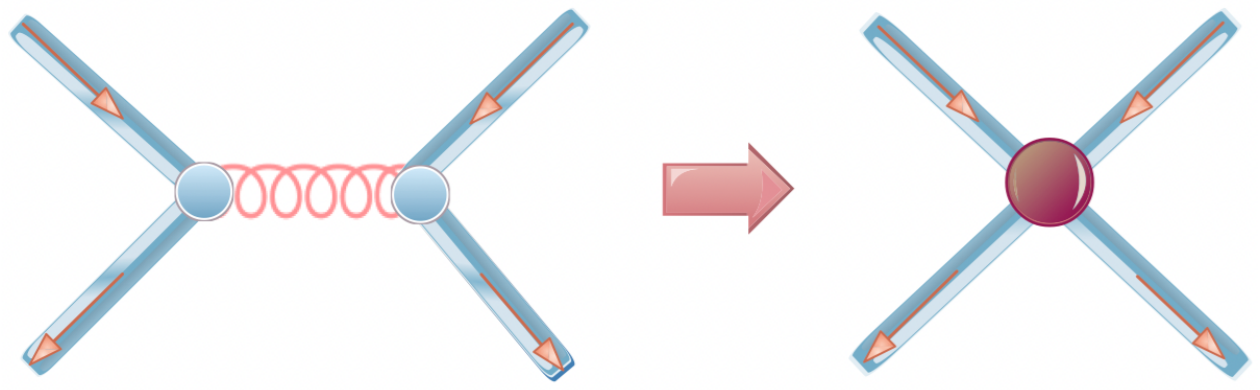}
    \vspace{-3.7cm}
    \caption{Diagrammatic representation of the CI model based on a momentum-independent gluon propagator employed in 
    Eq.~(\ref{eqn:contact_interaction}), yielding effective 4-point interaction between quarks. }
    \label{fig:ci}
\end{figure}

The starting point for our discussion is the dressed-quark propagator for a quark of flavor $f$,
which is obtained by solving the following SDE for the quark propagator or the so called gap equation\,:
\begin{align}
 S(p)^{-1} &= i\gamma\cdot p + m_{f} + \Sigma(p) \,,\nn \\
\Sigma(p) &= \frac{4}{3} \int \! \frac{d^4q}{(2\pi)^4} g^2 D_{\mu\nu}(p-q)
\gamma_\mu S(q) \Gamma_\nu(q,p) \,,  \label{gendse}
\end{align}
where $m_f$ is the current-quark mass in the Lagrangian, $D_{\mu\nu}(p)$ is
the gluon propagator and $\Gamma_\nu(q,p)$ is the quark-gluon vertex. 
 It is a well-established fact by now that the Landau gauge gluon propagator saturates in the infrared and a large effective mass scale is generated for the gluon, see for example Refs.~\cite{Boucaud:2011ug,Ayala:2012pb,Bashir:2013zha,Binosi:2016nme,Deur:2016tte,Rodriguez-Quintero:2018wma}. It also leads to the saturation of the effective strong coupling at large distances. This modern understanding of infrared QCD
forms the defining ideas of the CI~\cite{GutierrezGuerrero:2010md}.
We assume that the quarks interact, not through massless vector-boson exchange
but via a contact interaction.  Thus the gluon propagator no longer runs with a momentum scale but it is squeezed to a point in keeping with the infrared properties of QCD, see Fig~\ref{fig:ci}. 
Therefore, we can write\,: 
\begin{eqnarray}
\label{eqn:contact_interaction}
g^{2}D_{\mu \nu}(k)&=&4\pi\hat{\alpha}_{\mathrm{IR}}\delta_{\mu \nu} \, ,
\end{eqnarray}
\noindent  where $\hat{\alpha}_{\mathrm{IR}}=\alpha_{\mathrm{IR}}/m_g^2 \equiv \hat{\alpha}_{\mathrm{IRL}} $ for the light quarks $\tu,\td,\ts$. For heavier quarks, we choose  $ \hat{\alpha}_{\mathrm{IR}} =  \hat{\alpha}_{\mathrm{IRL}} / Z_H $. Increasing $Z_H$ for heavier quarks ensures a smaller effective coupling.  $m_g$ is interpreted as the infrared gluon mass scale generated dynamically within QCD \cite{Aguilar:2002tc,Bowman:2004jm,Aguilar:2006gr}. We take the current accepted value, i.e.,  $m_g=500\,\MeV$~\cite{Boucaud:2011ug,Aguilar:2017dco,Binosi:2017rwj,Gao:2017uox}. In the CI gap equation, the effective coupling which appears is $\hat{\alpha}_{\mathrm{IR}}$ instead of $\alpha_{\mathrm{IR}}$.  
 We choose $\alpha_{\rm IR}/\pi$ to be $0.36$ so that $\hat{\alpha}_{\mathrm{IR}}$ has exactly the same value as in all related previous works~\cite{Gutierrez-Guerrero:2010waf,Gutierrez-Guerrero:2019uwa,Gutierrez-Guerrero:2021rsx,Yin:2019bxe}.
The interaction vertex is bare, i.e.,
$\Gamma_\nu(q,p)=\gamma_\nu$.

This constitutes an algebraically simple but useful and predictive rainbow-ladder truncation of the SDE of the quark propagator whose
solution can readily be written as follows:
 \bea\label{DynamicalM}
 S(q,M_f) &\equiv &-i\gamma\cdot q \; \sigma_{V}(q,M_f)+\sigma_{S}(q,M_f)\,,
 \eea
 with
 \begin{eqnarray}
 \sigma_{V}(q,M_f)&=&\frac{1}{q^{2}+M_f^{2}}\, , \nn\\ 
 \sigma_{S}(q,M_f)&=& M_f
 \, \sigma_{V}(q,M_f) \,, 
 \end{eqnarray}
where $M_f$, in the CI, is momentum-independent dynamically generated dressed quark mass determined by
\begin{equation}
M_f = m_f + M_f\frac{4\hat{\alpha}_{\rm IR}}{3\pi}
\int_0^\infty \!ds \, s\, \frac{1}{s+M_f^2}\,\,. \label{gap-2}
\end{equation}
Our regularization procedure
 follows Ref.\,\cite{Ebert:1996vx}\,:
\begin{eqnarray}
\nonumber \frac{1}{s+M_f^2} & = & \int_0^\infty d\tau\,{\rm
e}^{-\tau (s+M_f^2)} \rightarrow \int_{\tau_{\rm UV}^2}^{\tau_{\rm
IR}^2} d\tau\,{\rm e}^{-\tau (s+M_f^2)}
\label{RegC}\\
& = & \frac{{\rm e}^{- (s+M_f^2)\tau_{\rm UV}^2}-e^{-(s+M_f^2)
\tau_{\rm IR}^2}}{s+M_f^2} \,, \label{ExplicitRS}
\end{eqnarray}
where $\tau_{\rm IR,UV}$ are, respectively, infrared and
ultraviolet regulators.  It is apparent from
Eq.\,(\ref{ExplicitRS}) that a finite value of $\tau_{\rm
IR}\equiv 1/\Lambda_{\rm IR}$ implements confinement by ensuring the
absence of quark production thresholds. Since Eq.\,(\ref{gap-2})
does not define a renormalisable theory, $\Lambda_{\rm
UV}\equiv 1/\tau_{\rm UV}$ cannot be removed but instead plays a
dynamical role, setting the scale of all mass dimensioned quantities.
Using Eq.\,\eqref{RegC}, the gap equation becomes,
 \bea \label{eq:gapeq}
  M_f = m_f + M_f \frac{4 \hat{\alpha}_{\rm IR}}{3 \pi }
  {\cal C}(M_f^2) \;,
 \eea
 where
 \bea
  \frac{{\cal C}(M^2)}{M^2} = \Gamma(-1,M^2 \tau_{\rm UV}^2) -
  \Gamma(-1,M^2 \tau_{\rm IR}^2) \; , 
 \eea
 and $\Gamma(\alpha,x)$ is the incomplete gamma-function.

 \begin{table}[t!]
 \caption{ \label{parameters} 
 The ultraviolet regulator and coupling constant for different quark combinations in mesons.  $\hat{\alpha}_{\mathrm {IR}}=\hat{\alpha}_{\mathrm{IRL}}/Z_H$, where $\hat{\alpha}_{\mathrm {IRL}}$ is extracted from the best-fit to data as explained in Ref.~\cite{Raya:2017ggu}.  $\Lambda_{\rm IR} = 0.24$ GeV is a fixed parameter.} 
\begin{center}
\label{parameters1}
\begin{tabular}{@{\extracolsep{0.0 cm}} | l | c | c | c |}
\hline \hline
 \, quarks \, &\,  $Z_{H}$ \, &\,  $\Lambda_{\mathrm {UV}}\,[\GeV] $ \,  &\,  $\hat{\alpha}_{\mathrm {IR}}$ \,
 \\
 \hline
 \rule{0ex}{2.5ex}
 $\, \tu,\td,\ts$ & 1 & 0.905 & \, 4.57 \,  \\ 
\rule{0ex}{2.5ex}
$\, \tc,\tu$ & \, 3.034 \, & 1.322 & 1.50 \\
\rule{0ex}{2.5ex}
$\, \tc,\ts$ & \, 8.172 \, & 0.584 & 0.56  \\
\rule{0ex}{2.5ex}
$\, \tc$     &  17.067 & 2.935 &  0.27\\
\rule{0ex}{2.5ex}
$\, \tb,\tu$ & 4.048 & 5.188 & 1.13 \\
\rule{0ex}{2.5ex}
 $\, \tb,\ts$ & 13.518 & 7.399 & 0.34 \\
\rule{0ex}{2.5ex}
$\, \tb,\tc$   & 30.972 & 7.610 &\,   0.15\, \\
\rule{0ex}{2.5ex}
$\, \tb$     & \, 130.585 \,  & 14.958 & 0.03   \\
\hline \hline
\end{tabular}
\end{center}
\end{table}
 We report results for vectors mesons  using the parameter values listed in Tables~\ref{parameters} and~\ref{table-M}, whose variation with quark mass was dubbed as 
 {\em heavy parameters} in Ref.~\cite{Gutierrez-Guerrero:2019uwa}. 
 In this approach, the coupling constant and the ultraviolet regulator vary as a function of the quark mass. This behavior was first suggested  in Ref.~\cite{Bedolla:2015mpa} and later adopted in several subsequent works~\cite{Bedolla:2016yxq,Raya:2017ggu,Gutierrez-Guerrero:2019uwa,Yin:2019bxe,Yin:2021uom}. Table~\ref{table-M} presents the current quark masses $m_f$ used herein and the dynamically generated dressed masses $M_f$ of $\tu$, $\ts$, $\tc$ and $\tb$ computed from the gap equation, Eq.~(\ref{eq:gapeq})\footnote{We assume isospin symmetry all along this work.}.
%

\begin{table}[t!]
\caption{\label{table-M}
Current ($m_{f}$) and dressed masses
($M_{f}$) for quarks in GeV, required as an input for the BSE and the EFFs.}
\vspace{0.3cm}
\begin{tabular}{@{\extracolsep{0.0 cm}} | c | c | c | c | }
\hline 
\hline
 $m_{\tu}=0.007$ &$m_{\ts}=0.17$ & $m_{\tc}=1.08$ & $m_{\tb}=3.92$   \\
 \rule{0ex}{2.5ex}
 $ M_{\tu}=0.367$ \, & \, $  M_{\ts}=0.53$\; \, &\,   $  M_{\tc}=1.52$ \, &\,  $  M_{\tb}=4.75$   \\
 \hline
 \hline
\end{tabular}
\end{table}

A meson can consist of heavy ($Q$) or light ($q$) quarks. Therefore, we present the study of all heavy ($Q\bar{Q}$), heavy-light ($Q\bar{q}$) and (review) light ($q\bar{q}$) V-mesons.
We commence by setting up the BSE for mesons by employing a kernel which is consistent with theoretical constraints, i.e., the gap equation must obey the AV Ward-Takahashi identity and low energy Goldberger-Treiman relations, see Ref.~\cite{Gutierrez-Guerrero:2010waf} for details. 
The V mesons are $J^{PC}=1^{--}$ states. 
The solution of the BSE yields BSAs whose general form depends not only on the spin and the parity of the meson under consideration  but also on the interaction kernel employed as explained in the next sub-section.

\subsection{The Bethe-Salpeter equation: meson masses}
\label{bse-a}
\begin{figure}[b!]
   \centering
    \includegraphics[scale=0.5,angle=0]{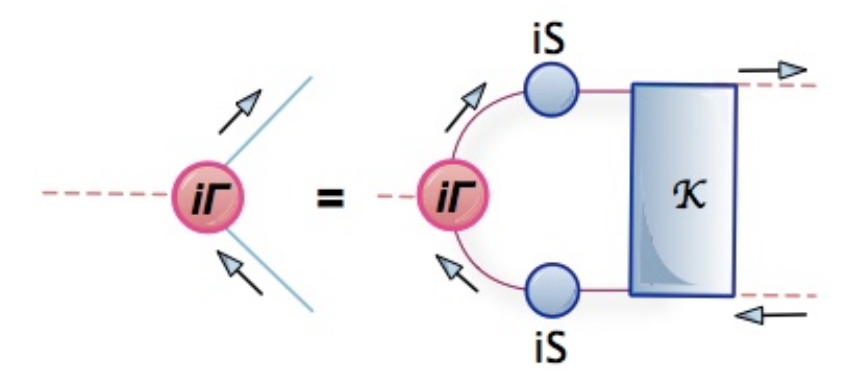}
    \caption{Diagrammatic representation of the BSE. Blue (solid) circles represent dressed quark propagators $S$, red (solid) circle corresponds to the meson BSA $\Gamma$ while the blue (solid) rectangle is the dressed-quark-antiquark interaction kernel ${\mathcal {K}}$.}
    \label{fig:BSEfig}
\end{figure}
  The relativistic bound-state problem for hadrons characterized by two valence quarks can be studied by using the
 homogeneous BSE whose diagrammatic representation can be seen in  Fig.~\ref{fig:BSEfig}. This equation is mathematically expressed  as~\cite{Salpeter:1951sz},
 \begin{equation}
[\Gamma(k;P)]_{tu} = \int \! \frac{d^4q}{(2\pi)^4} [\chi(q;P)]_{sr} {\mathcal K}_{tu}^{rs}(q,k;P)\,,
\label{genbse}
\end{equation}
where $[\Gamma(k;P)]_{tu}$ represents the bound-state's BSA and $\chi(q;P) = S(q+P)\Gamma S(q)$ is the BS wave-function; $r,s,t,u$ represent colour, flavor and spinor indices; and ${\mathcal K}$ is the relevant quark-antiquark scattering kernel. This equation possesses solutions on that discrete set of $P^2$-values from which bound-states exist.
We enlist the masses and BSAs of V mesons in Table \ref{par-AllFF}.
\begin{table}[htb]
\caption{\label{par-AllFF} Calculated values of the BSAs and the masses of the V mesons in the CI model using the parameters in Tables~\ref{parameters} and~\ref{table-M} (compare the parameters with  the ones in Ref.~\cite{Gutierrez-Guerrero:2019uwa}). Entries marked as * are not experimental results. These values were obtained with lattice QCD methods.} 
\begin{tabular}{@{\extracolsep{0.1 cm}} | c | cc | c | c | }
\toprule
 \rule{0ex}{2.5ex}
 &  Mass[GeV]  & $E_{\Me}$ & $m_{\Me}^{\rm exp}$[GeV] & error [\%]  \\ 
 \hline
 \rule{0ex}{2.5ex}
 $\rho(\tu\bar\td)$ \, & 0.93 & 1.53 \, & 0.78 & 19.23 \\
 \rule{0ex}{2.5ex}
 $K_1(\tu\bar\ts)$\, & 1.03 & 1.63 \, & 0.89  & 15.73  \\
\rule{0ex}{2.5ex}
$\phi(\ts\bar{\ts})$ \, & 1.13 & 1.74\, & 1.02 &   10.78 \\
\rule{0ex}{2.5ex}
$D^{*0}(\tc\bar{\tu})$\, & 2.05 & 1.23 \,& 2.01 &     1.99 \\
\rule{0ex}{2.5ex}
$D^*_{\ts}(\tc\bar{\ts})$\, & 2.30 & 0.55 \,& 2.11 &     9.00 \\
\rule{0ex}{2.5ex}
$B^{+*}(\tu\bar{\tb})$ \, &  4.93 & 2.73 \,& 5.33 &  7.50 \\
\rule{0ex}{2.5ex}
$B^{0*}_{\ts}(\ts\bar{\tb})$\, & 5.39 & 1.17 \,& 5.42 &   0.55 \\
\rule{0ex}{2.5ex}
$B^{0*}_{\tc}(\tc\bar{\tb})$\, & 6.25 & 1.06 \,& 6.33* &   1.26 \\
\rule{0ex}{2.5ex}
$\Jpsi(\tc\bar{\tc})$ \, & 3.15 & 0.51 \,& 3.10  & 1.61 \\
\rule{0ex}{2.5ex}
$\Upsilon(\tb\bar{\tb})$ \, & 9.51 &  0.48 \, & 9.46 &  0.53 \\
\hline
\hline
\end{tabular}
\end{table}
A general decomposition of the BSA for the V mesons ($\fd\fu$)  in the CI model has the following form,
 \bea \label{bse-v-av}  &&
\Gamma_{\Me,\mu} = \gamma_\mu^\perp E_{\Me}(P)\,,
\eea
  where $E_V(P)$ is the BSA, $P$ is the total meson momentum and  \bea \gamma_{\mu}^{T}=\gamma_{\mu}-\frac{ \gamma \cdot P \;
 }{P^{2}} \, P_{\mu}  \,.
 \eea
 After this initial and required set up of the gap equation and the BSE, we now turn our attention to the description of the quark-photon interaction vertex.

 \subsection{The quark-photon vertex}
The quark-photon vertex, denoted by   $\Gamma_{\mu}^{\gamma}(k_+,k_-,M_{\fd})$, and studied in great detail in the literature,~\cite{Guzman:2023hzq,Lessa:2022wqc,Albino:2018ncl,Bashir:2011vg,Bashir:2007qq,Bashir:2004hh,Bashir:1999bd,Bashir:1995qr}, is related to the quark propagator through the following 
V Ward-Takahashi
identity:
 \bea
  i P_{\mu} \Gamma_{\mu}^{\gamma}(k_+,k_-,M_{\fd}) =
  S^{-1}(k_+,M_{\fd}) - S^{-1}(k_-,M_{\fd}) \,. \nonumber \\ \label{VWTI}
 \eea
 This identity is crucial for a sensible study of a bound-state's EFF. It is determined through the following inhomogeneous BSE,
 \begin{align}
 \Gamma_{\mu}^{\gamma}= \gamma_{\mu} - \frac{16 \pi \hat{\alpha}_{\rm IR}}{3} 
  \int  \frac{d^4q}{(2 \pi)^4} \gamma_{\alpha} \chi_{\mu}(q_+,q,M_{\fd})
 \gamma_{\alpha} \, ,\label{eqvertex}
 \end{align}
where $\chi_{\mu}(q_+,q,M_{\fd})  =  S(q+P,M_{\fd}) \Gamma_{\mu}(Q)S(q,M_{\fd})$. Note that we have omitted explicit functional dependence of $\Gamma_\mu^\gamma=\Gamma_\mu^\gamma(Q,M_{\fd})$ for notational simplicity.
Owing to the momentum-independent nature of the interaction
kernel, the general form of the solution is
  \bea
 && \hspace{-4mm} \Gamma_{\mu}^{\gamma}= \gamma_{\mu}^{L}(Q)P_{L}(Q^{2},M_{\fd}) +
 \gamma_{\mu}^{T}(Q)P_{T}(Q^{2},M_{\fd}), 
 \eea
 where $\gamma_{\mu}^{L} + \gamma_{\mu}^{T} = \gamma_{\mu}$. 
 Inserting this general relation into Eq.~(\ref{eqvertex}), one readily
 obtains
 \bea
 \hspace{-2mm} P_{L} =1 \,, \quad 
 P_T= \frac{1}{1+K_\gamma(Q^2,M_{\fd})} \,, \label{PTQ2}
 \eea 
\vspace{-0.8cm}
\begin{figure}[htbp]
\centering
\includegraphics[width=9cm]{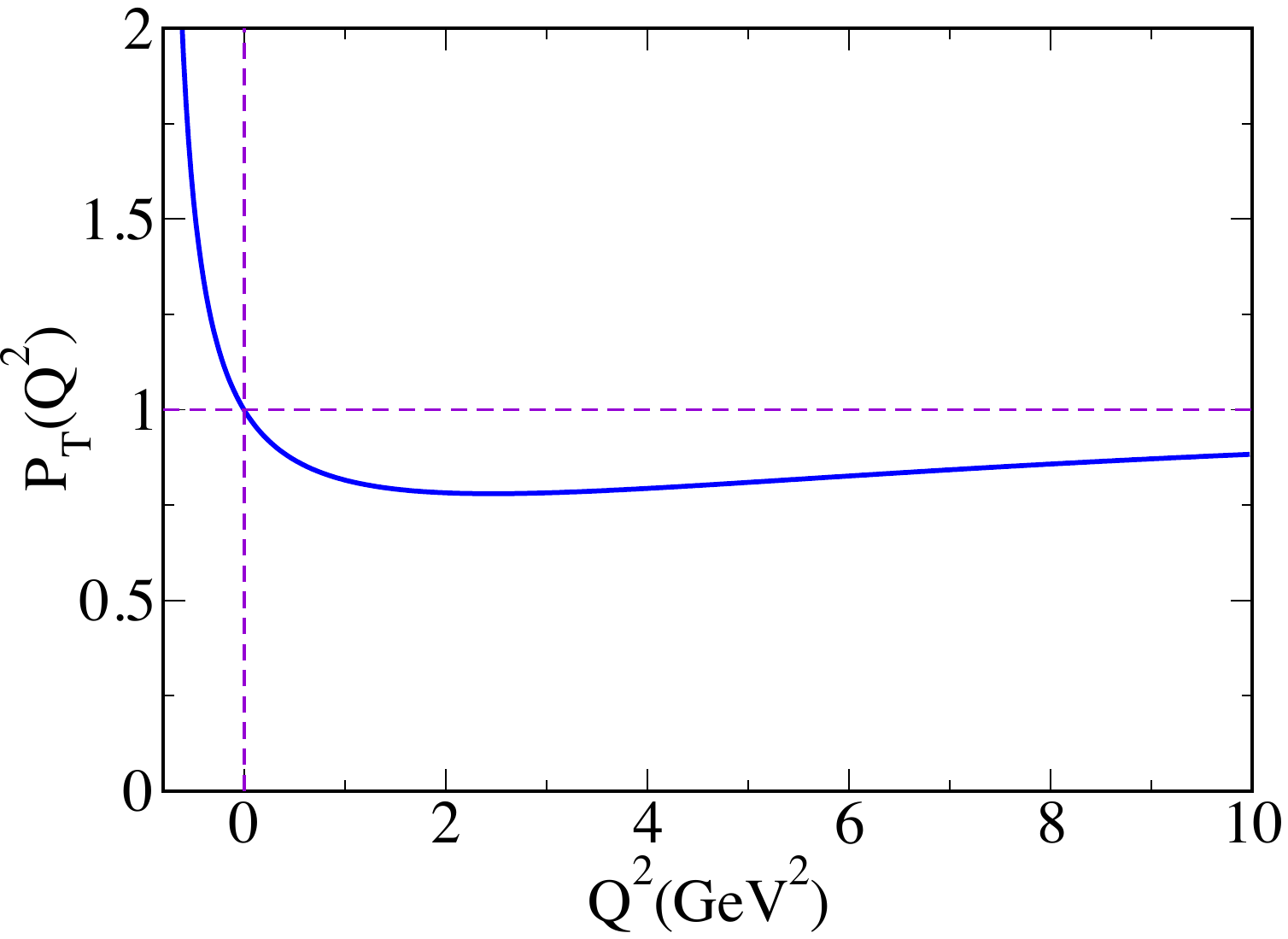}
\caption{Dressing function of the transverse quark-photon vertex, $P_T(Q^2)$, defined in~\protect{Eq.~(\ref{PTQ2}).}}
\label{fig:rhopole}
\end{figure}

\noindent with
\bea
&& \hspace{-1cm} K_\gamma(Q^2,M_{\fd}) = \frac{4 \hat{\alpha}_{\rm
IR}}{3\pi} 
\int_0^1d\alpha\, \alpha(1-\alpha) Q^2\,\bar{\mathcal{C}}_1(\omega)
\,, \\
&& \bar{\cal C}_1(z) = - \frac{d}{dz}{\cal
C}(z)= \Gamma(0,z\,  \tau_{\rm UV}^2)-\Gamma(0,z \,
\tau_{\rm IR}^2)\,, \\
&& \hspace{0.55cm} \omega=\omega(M_{\fd}^2,\alpha,Q^2) 
 =M_{\fd}^2 + \alpha(1-\alpha) Q^2 \,.
\eea
One can clearly observe from Fig.~\ref{fig:rhopole} that $P_{T}(Q^{2})
\rightarrow 1$ when $Q^2 \rightarrow \infty$, yielding the
perturbative bare vertex $\gamma_{\mu}$ as expected.
This quark-photon vertex provides us with the required electromagnetic interaction capable of probing the FFs of mesons through a triangle diagram which keeps the identity of the meson bound state intact.

\section{\label{sec:eff} Electromagnetic Form Factors}
After defining all the ingredients and identifying the tools necessary for our purpose in the previous section, we now describe the details of the triangle diagram calculation of the EFFs in this section. 
\subsection{The general $M\gamma M$-vertex}
\begin{figure}[htt]
\centerline{
\includegraphics[scale=0.25,angle=0]{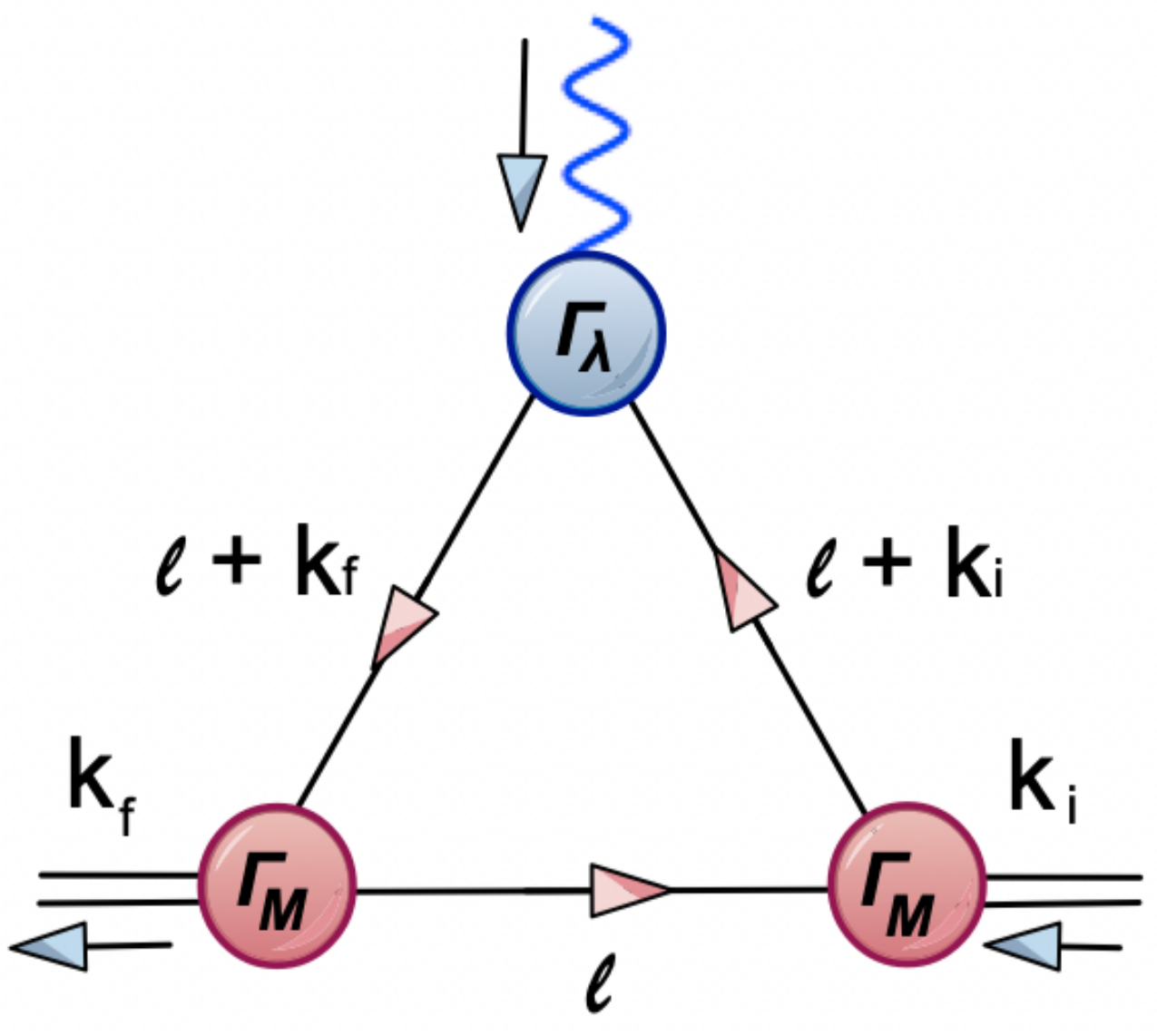}
}
    \caption{\label{vertex-1} The triangle diagram for the impulse approximation to the $M\gamma M$ vertex.}
\end{figure}

Let us start from the general considerations for the elastic electromagnetic interaction of a V meson, $M$. In the impulse approximation, the $M \gamma M$-vertex, which describes the elastic interaction between the meson $M(\fd\fu$) and a photon, reads
\bea \Lambda^{M,\fd}_{\lambda\mu\nu}&=&N_c\int \frac{d^{4}\ell}{(2\pi)^{4}}
 {\rm Tr}\;\mathcal{G}_{\lambda\mu\nu}^{M,\fd} \,,
 \eea
where
 \bea
 \nonumber\mathcal{G}^{M,\fd}_{\lambda\mu\nu}&=& 
  \, i\Gamma^{M}_{\mu}(k_{f}) \, S(\ell+k_{i},M_{\fd}) \, i\Gamma_{\lambda}(Q,M_{\fd})\\
  \nn &\times& 
  \, S(\ell+k_{f},M_{\fd})
 i\bar{\Gamma}^{M}_{\nu}(-k_{i}) \, S(\ell,M_{\fu}) \,. \label{General-FF}
 \eea
$\Gamma^M_{\mu}$ are the BSAs for V mesons given in Eqs.~(\ref{bse-v-av}), $S(p,m)$ is the quark propagator and $M_f$ is the dressed quark mass in Eq.~(\ref{eq:gapeq}).
The notation assumes that it is the quark $\fd$ which interacts with the photon while the antiquark $\fu$ remains a spectator. We can define 
$\Lambda^{M,\fu}_{\lambda\mu\nu}$ similarly. 
 Furthermore, we denote the incoming photon momentum by $Q$ while the incoming and outgoing momenta of $M$ by:
$k_{i}=k-Q/2$ and $k_f=k+Q/2$, respectively.
The assignments of momenta are shown in the triangle diagram of Fig.~\ref{vertex-1}.
$\Lambda^{M,\f}_{\lambda\mu\nu}$ contains complete information of the EFFs of different mesons under study. The contribution arising from the interaction of the photon with quark $\fd$ can be represented as $F^{M,\fd} (Q^2)$ (stemming from $\Lambda^{M,\fd}_{\lambda\mu\nu}$) while the contribution due to its interaction with antiquark $\fu$ can be represented as $F^{M,\fu} (Q^2)$ (coming from $\Lambda^{M,\fu}_{\lambda\mu\nu}$). The total EFF $F^M (Q^2)$ is defined as~\cite{Hutauruk:2016sug}\,:
\begin{equation}\label{eqn:TotalMesonFF}
F^M (Q^2) = e_{\fd} F^{M,\fd} (Q^2) + e_{\fu} F^{M,\fu} (Q^2)\,,
\end{equation}
where $e_{\fd}$ and $e_{\fu}$ are the quark and the antiquark electric charges, respectively~\footnote{For neutral mesons composed of same flavored quarks, the total EFF is simply defined as $F^{M} = F^{M,\fd}$.}.
\noindent
All information necessary for
the calculation of the EFFs is now complete. We can employ numerical values of the parameters listed in Tables~\ref{parameters} and~\ref{table-M} to compute the EFFs. Our evaluated analytical expressions and numerical results for V mesons occupy the details of the following subsection. 

\subsection{\label{VFF} The vector meson elastic form factors}

\begin{figure*}[t!]
\begin{tabular}{@{\extracolsep{-2.3 cm}}cc}
\hspace{-2cm}
\includegraphics[scale=0.65]{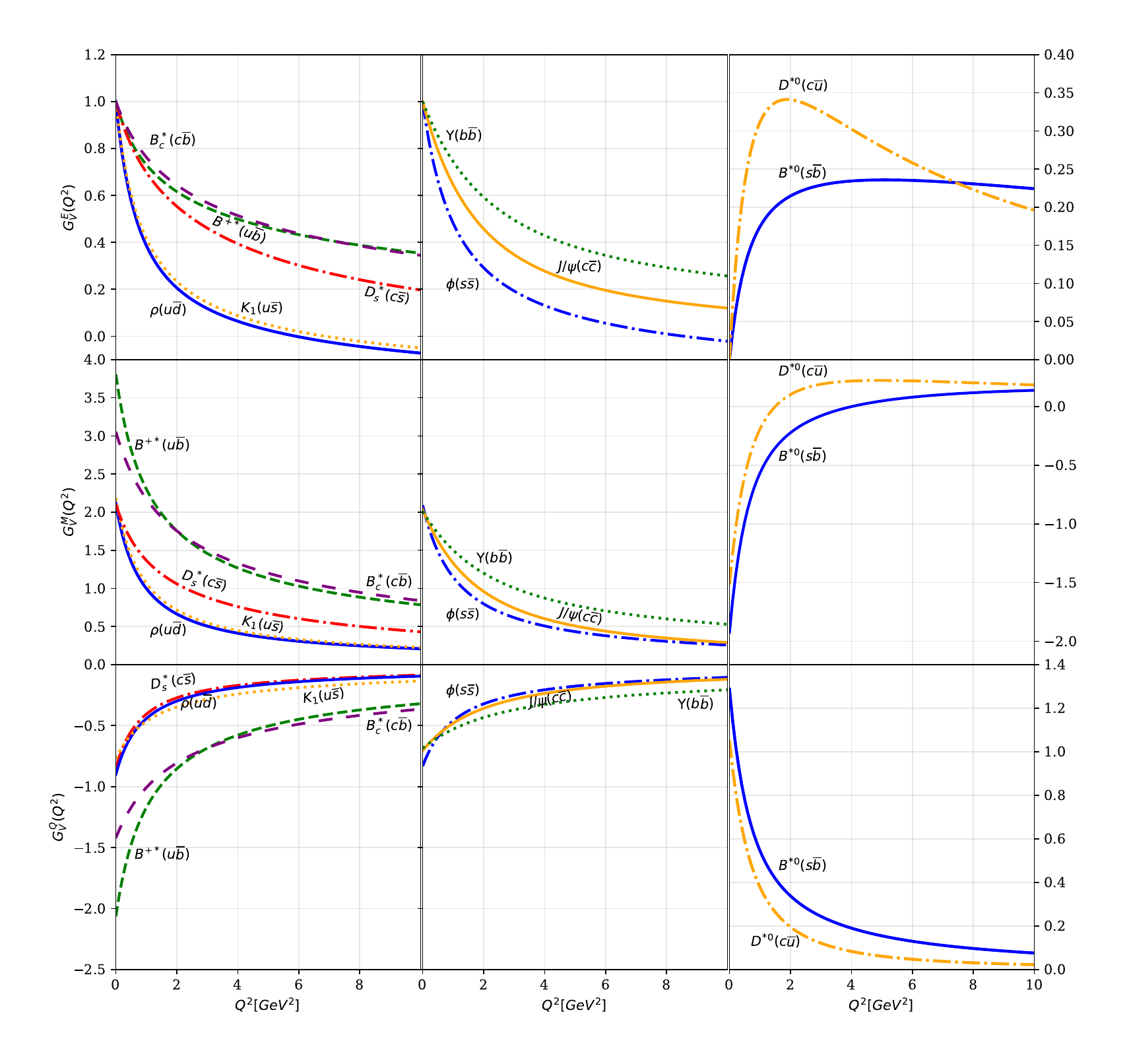}
\end{tabular} 
\vspace{-1cm}
\caption{V mesons' electric, magnetic, and quadrupole FFs are displayed in the top, middle, and bottom rows, respectively. FFs are depicted on the left, center, and right columns for the charged V mesons, neutral V mesons consisting of identical flavored quarks and the neutral V mesons composed of different flavored quarks, respectively. }\label{plotVE}
\end{figure*}
 According to quantum mechanics and the quark model, the V meson may be viewed as the valence-quark spin-flip partner of the PS mesons.
Consequently, the attributes of these two states differ noticeably. One of the most important fact is that the latter are distinguished by having only one EFF, while the former are characterized by three of them. For the V mesons, the extraction of $F_j^{M,f_1}$ requires appropriate projections since due to the spin $1$ nature of these particles, the corresponding relation with 
$\Lambda^{M,\fd}_{\lambda\mu\nu}$ is more elaborate\,:  
\begin{align}\label{Eq:VFF}
     \Lambda_{\lambda \mu \nu}^{\Me,\fd} = \sum_{j=1}^3 T_{\lambda \mu \nu}^{(j)}(k,Q) \, F_j^{\Me,\fd}(Q^2) \, ,
 \end{align}with the tensors $T_{\lambda \mu \nu}^{(j)}$ expressed as\,:
%
%
 \begin{align}
 T_{\lambda \mu \nu}^{(1)}(k,Q) & =  2 k_\lambda\, {\cal
 P}^T_{\mu\alpha}(k_i) \, {\cal P}^T_{\alpha\nu}(k_f)\,, \\
 T_{\lambda \mu \nu}^{(2)}(k,Q) & =  \left[Q_\mu - k_{i\mu} \frac{Q^2}{2 M_{M}^2}\right] {\cal P}^T_{\lambda\nu}(k_f) \nonumber \\
 &- \left[Q_\nu + k^f_\nu \frac{Q^2}{2 M_{M}^2}\right] {\cal P}^T_{\lambda\mu}(k_i)\,, \\
 T_{\lambda\mu \nu}^{(3)}(k,Q) & =  \frac{k_\lambda}{M_{M}^2}\, \left[Q_\mu - k_{i\mu} \frac{Q^2}{2 M_{M}^2}\right] \nonumber \\
 &\times \left[Q_\nu + k_{f\nu} \frac{Q^2}{2 M_{M}^2}\right] \,.
 \end{align}
The transverse projector in the above relations is given by the following expression\,:
\begin{eqnarray} {\cal P}_{\alpha
\beta}^T(P) = \delta_{\alpha \beta} - {P_{\alpha}
P_{\beta}}/{P^2} \,.
\end{eqnarray}
%
After introducing the Feynman parameters and integrating over the internal momentum $l$, the EFFs defined in Eq.~(\ref{Eq:VFF}), $F_i^{V,f_1}$, can be projected out to be\,: 
\begin{eqnarray}
\label{eqn:vecmes}
\nn && \hspace{-1cm} F^{\Me,\fd}_i(Q^2) = \frac{3}{4\pi^2} E^2_{\Me}  \, P_T(Q^2)
\int_0^1 d\alpha \, d\beta \; \alpha \, \\ 
&& \hspace{7mm}  \times \bigg[
\mathcal{A}^{\Me}_i \, \overline{\mathcal{C}}_1(\omega_1)  +(\mathcal{B}^{\Me}_i -\mathcal{A}^{\Me}_i \, \omega_2) \, \overline{\mathcal{C}}_2(\omega_2) 
\bigg] \, ,
\end{eqnarray}
where the labels $i=1,2,3$ correspond to the three EFFs. The analytic expressions for $\mathcal{A}_i^V$ and $\mathcal{B}_i^V$ can be found in the Appendix~\ref{App:EM}. For V mesons, the definition given in 
Eq.~(\ref{eqn:vecmes}) is important to define the three experimentally accessible and physically suggestive EFFs: electric, magnetic and quadrupole form factors, defined as
 \begin{subequations}
\begin{align}
G_E^{V}(Q^2) & =  F_1^{ V}(Q^2)+\frac{2}{3} \eta \, G_Q^{ V}(Q^2)\,,\\
G_M^{ V}(Q^2) & =  - F_2^{ V}(Q^2)\,, \label{Definemu}\\
%
G_Q^{ V}(Q^2) & =  F_1^{ V}(Q^2) + F_2^{ V}(Q^2) + \left[1+\eta\right] F_3^{ V}(Q^2)\,,
\end{align}
\end{subequations}
where $\eta=Q^2/(4 M_{ M}^2)$ and $M_M$ is the mass of the V meson. In the limit $Q^2\to 0$, these EFFs define the charge, magnetic ($\mu_V$) and quadrupole moments ($\mathcal{Q}_V$) of the V meson; viz.,
\begin{subequations}
\begin{eqnarray}
\label{chargenorm}
G_E^{ V}(Q^2=0) & = & 1\,, \\
G_M^{ V}(Q^2=0) & = & \mu_{ V},\;
G_Q^{ V}(Q^2=0) = \mathcal{Q}_{ V}\,.
\end{eqnarray}
\end{subequations}
EFFs of light, heavy-light, and the heaviest V mesons are displayed in Fig.\,\ref{plotVE}. The asymptotic limit of QCD reported in Ref.~\cite{Brodsky:1992px} predicts 
\bea
  G_{E}(Q^2):G_{M}(Q^2):G_Q(Q^2) \stackrel{Q^2 \rightarrow \infty}{=}
  1 - \frac{2}{3} \eta : 2 : -1 \,.
 \eea
Although none of these predictions are strictly satisfied in our computation, as noted in Ref.~\cite{Roberts:2011wy,Raya:2017ggu}, this relation is recovered for $\Lambda_{\rm UV} \rightarrow \infty$ but at the cost of a logarithmic divergence in the individual EFFs. Therefore, we conclude that a vector-vector CI cannot reasonably be regularized in a manner fully consistent with the constraints of asymptotic QCD. A more involved model must be constructed to recover the predictions of Ref.~\cite{Brodsky:1992px} exactly.
As has been seen in several works including Ref.~\cite{Roberts:2011wy}, the electric FF of the $\rho$-meson might have a zero crossing, a behavior not found for S and PS mesons~\cite{Hernandez-Pinto:2023yin}. 
Similarly, we find that the electric FF of mesons composed of light quarks, the $K_1$ and the $\phi$ mesons, present a zero crosssing, {\em albeit} at increasing values of momentum transfer, $Q^2\simeq7,10$ GeV$^2$, respectively. However, in the case of mesons composed by at least one heavier quark, $c$ or $b$, we do not observe any zero crossing even for large values of momentum transfer $Q^2\simeq50$ GeV$^2$, a similar behavior was observed in Ref.~\cite{Raya:2017ggu} within the CI for $J/\psi$ and $\Upsilon$ mesons.

We can now readily compute the charge, magnetic and quadrupole radii, using 
\begin{equation}\label{fradii}
 r_{i}^2 =
-6\left.\frac{\mathrm{d}G_{i}(Q^{2})}{\mathrm{d}Q^{2}}\right|_{Q^{2}=0}
\,,
\end{equation}
with $i\in \{E,M,Q\}$. It is important to remark that positive slopes shall appear in some EFFs; in those cases, the definition for charge radii is opted to remove the minus sign in the equation to avoid imaginary numbers. Since the properties of V mesons are also under investigation by several other groups, we present a comparison of charge radii with the NJL model~\cite{Luan:2015goa}, with the method based on the Bag model \cite{Simonis:2016pnh} and in a symmetry-preserving approach to the two valence-body continuum bound-state problem \cite{Bhagwat:2006pu} in Tabs.~\ref{fbo},~\ref{fbo1} and~\ref{fbo2}. Experimental data on these static properties are absent to date. Therefore, results from other theoretical models are included in the Tables. 
\begin{table}[t!]
\begin{center}
\caption{\label{fbo}  Electric, magnetic
and quadrupole radii (in fm)  for the lightest mesons composed of quarks $\tu,\td, \ts$. Our results are compared with the NJL model \cite{Luan:2015goa}, method based on the Bag model \cite{Simonis:2016pnh}, and a symmetry-preserving approach to the two valence-body continuum bound-state problem \cite{Bhagwat:2006pu}.
For the $\rho(\tu\bar{\td})$-meson one of the first works on vector $\times$ vector CI \cite{Roberts:2011wy} predicted $r^{E}=0.56$, $r^{M}=0.51$, $r^{Q}=0.51$, $\mu=2.11$ and $\mathcal{Q}=-0.85$. 
A lattice QCD simulation~\cite{Owen:2015gva} yields $r^E = 0.82$ fm, and $\mu=2.21$; 
 } 
\vspace{0 cm}
\begin{tabular}{|c@{\extracolsep{-0.6 cm}}c@{\extracolsep{0.2 cm}}ccccc|}
\hline
\hline
\rule{0ex}{3.0ex}
& & $r^{E}$ & $r^{M}$ & $r^{Q}$ & $\mu$ & $\mathcal{Q}$  \\ 
\hline
 \rule{0ex}{3.0ex}
 \multirow{2}{2.0cm}{$\rho(\tu\bar{\td})$ } & Our& 0.560 & 0.745 & 0.472 & 2.11 & $-0.85$
\\ 
\rule{0ex}{3.0ex}
& Ref. \cite{Bhagwat:2006pu}  & 0.54 &- &-&2.01  & $-0.41$ \\
\rule{0ex}{3.0ex}
& Ref. \cite{Xu:2019ilh} & 0.72  &- &-& 2.01  & $-0.36$ \\
\rule{0ex}{3.0ex}
 & Ref. \cite{Owen:2015gva}  &0.82 &- & - &2.07 & $-$0.70     \\
 \rule{0ex}{3.0ex}
 & Ref. \cite{Shultz:2015pfa} &0.55 &- & - &2.17 & $-$0.55    \\
\rule{0ex}{3.0ex}
 & Ref. \cite{Gurjar:2024wpq}  &0.75 &- & - &2.40 & $-$0.42    \\
 \rule{0ex}{3.0ex}
& Ref. \cite{Luan:2015goa}  & 1.12 &- &- &2.54 &- \\
\rule{0ex}{3.0ex}
& Ref. \cite{Simonis:2016pnh}  &-  &-& -& 2.10 &- \\
\rule{0ex}{3.0ex}
& Ref. \cite{Badalian:2012ft}  &- &- &- & 2.37  &-  \\
\rule{0ex}{3.0ex}
 & Ref. \cite{Qian:2020utg}  &0.44 &- & - &2.15 & $-$0.98     \\
\rule{0ex}{3.0ex}
 & Ref. \cite{Zhong:2023cyc}  &0.52 &- & - &1.92 & $-$0.44    \\
\rule{0ex}{3.0ex}
 & Ref. \cite{Carrillo-Serrano:2015uca}  &0.82 &- & - &2.48 & $-$1.09     \\
\rule{0ex}{3.0ex}
 & Ref. \cite{Xu:2024fun}  &0.72 &- & - &2.01 & $-$0.36     \\
\hline
 \rule{0ex}{3.0ex}
 \multirow{2}{2.0cm}{$K_1(\tu\bar{\ts})$ } & Our  & 0.535 & 0.733 & 0.471 & 2.18 & $-0.90$ 
\\ 
\rule{0ex}{3.0ex}
& Ref. \cite{Bhagwat:2006pu}  & 0.43 &- &-&2.23  & $-0.38$ \\
\rule{0ex}{3.0ex}
& Ref. \cite{Xu:2019ilh} &0.64 &- &- & 2.22  & $-0.31$ \\
\rule{0ex}{3.0ex}
& Ref. \cite{Luan:2015goa} & 1.05 & - & - & 2.26 & -
\\
\rule{0ex}{3.0ex}
& Ref. \cite{Simonis:2016pnh}  &  -& - & - & 2.06 & - \\
\rule{0ex}{3.0ex}
& Ref. \cite{Badalian:2012ft} &- &- &- &2.19 &-  \\
\rule{0ex}{3.0ex}
 & Ref. \cite{Xu:2024fun}  &0.68 &- & - &2.12 & $-$0.43     \\
\hline
 \rule{0ex}{3.0ex}
 \multirow{1}{2.0cm}{$\phi(\ts\bar{\ts})$ } & Our  & 0.472 & 0.632 & 0.396 & 2.09 & $-0.83$
\\ 
\rule{0ex}{3.0ex}
& Ref. \cite{Xu:2019ilh} &0.52 &- &- & 2.08  & $-0.32$ \\
\hline
\hline
\end{tabular}
\end{center}
\end{table}
\begin{table}[t!]
\begin{center}
\caption{ \label{fbo1}  Electric, magnetic
and quadrupole radii (in fm) for the heavy-light mesons. Our results are compared with the NJL model \cite{Luan:2015goa}, method based on the bag model \cite{Simonis:2016pnh}, and a symmetry-preserving approach to the two valence-body continuum bound-state problem \cite{Bhagwat:2006pu}.     
 } 
\vspace{0 cm}
\begin{tabular}{|c@{\extracolsep{-0.6 cm}}c@{\extracolsep{0.2 cm}}ccccc|}
\hline
\hline
\rule{0ex}{3.0ex}
& & $r^{E}$ & $r^{M}$ & $r^{Q}$ & $\mu$ & $\mathcal{Q}$  \\ 
\hline
 \rule{0ex}{3.0ex}
 \multirow{2}{2.0cm}{$D^{*0}(\tc\bar{\tu})$ } &  Our &  0.424 & 0.820 & 0.572 & $-1.51$ & 1.05 
\\ \rule{0ex}{3.0ex}
& Ref. \cite{Luan:2015goa} & 0.72 &- &- & 1.16 & -
\\ 
\rule{0ex}{3.0ex}
& Ref. \cite{Simonis:2016pnh}  & - &- &- & $-1.21$ &- \\
\rule{0ex}{3.0ex}
 & Ref. \cite{Xu:2024fun}  &0.60 &- & - &$-$2.30 & 0.76     \\
 \rule{0ex}{3.0ex}
& Ref. \cite{Aliev:2019lsd} &- &- &- & 0.30  & $0.25$ \\
\hline
 \rule{0ex}{3.0ex}
 \multirow{2}{2.0cm}{$D_{\ts}^*(\tc\bar{\ts}$) } & Our & 0.369 & 0.588 & 0.361 & 2.10 & $-0.71$
\\ \rule{0ex}{3.0ex}
& Ref. \cite{Luan:2015goa}  & 0.49 & - &- & 0.98 & -
\\ 
\rule{0ex}{3.0ex}
& Ref. \cite{Simonis:2016pnh}  & - &- &- & 0.87 &- \\
\rule{0ex}{3.0ex}
 & Ref. \cite{Xu:2024fun}  &0.38 &- & - &2.27 & $-$0.46     \\
 \rule{0ex}{3.0ex}
& Ref. \cite{Aliev:2019lsd} &- &- &- & 1.00  & $-0.60$ \\
\hline
 \rule{0ex}{3.0ex}
 \multirow{2}{2.0cm}{$B^{+*}(\tu\bar{\tb})$ } & Our  & 0.335 & 0.818 & 0.632 & 3.80 & $-2.06$
\\ \rule{0ex}{3.0ex}
& Ref. \cite{Luan:2015goa}  & 0.96  &- &- & 1.47 &- 
\\ 
\rule{0ex}{3.0ex}
& Ref. \cite{Simonis:2016pnh}  & - &- &- & 1.47 &- \\
\rule{0ex}{3.0ex}
 & Ref. \cite{Xu:2024fun}  &0.66 &- & - &7.88 & $-$2.17     \\
 \rule{0ex}{3.0ex}
& Ref. \cite{Aliev:2019lsd} &- &- &- & 0.90  & $-0.80$ \\
\hline
 \rule{0ex}{3.0ex}
 \multirow{2}{2.0cm}{$B_{\ts}^{0*}(\ts\bar{\tb})$ } & Our  & 0.300 & 0.825 & 0.596 & $-1.82$ & 1.20
\\ 
\rule{0ex}{3.0ex}
& Ref. \cite{Simonis:2016pnh}  &-  &- &- & $-0.48$ &- \\
\rule{0ex}{3.0ex}
 & Ref. \cite{Xu:2024fun}  &0.35 &- & - &$-$2.77 & 0.89     \\
\rule{0ex}{3.0ex}
& Ref. \cite{Aliev:2019lsd} &- &- &- & -0.17  & $-0.17$ \\
\hline
\hline
\end{tabular}
\end{center}
\end{table}
\begin{table}[h!]
\begin{center}
\caption{ \label{fbo2}  Electric, magnetic, and quadrupole radii (in fm)  for the heaviest mesons compared with the NJL 
model~\cite{Luan:2015goa}, method based on Bag model~\cite{Simonis:2016pnh}, a symmetry-preserving approach to the two valence-body continuum bound-state problem \cite{Bhagwat:2006pu}.
Using the same treatment shown in Ref.~\cite{Raya:2017ggu}, but different parameters, the results for $J/\Psi(\tc\bar{\tc})$ are $r^{E}=0.262$, $r^{M}=0.254$, $r^{Q}=0.240$, $\mu=2.047$ and $\mathcal{Q}=-0.748$ and for $\Upsilon(\tb\bar{\tb})$-meson the results are  $r^{E}=0.197$, $r^{M}=0.195$, $r^{Q}=0.182$, $\mu=2.012$ and $\mathcal{Q}=-0.704$. In lattice QCD~\cite{Dudek:2006ej} $J/\Psi(\tc\bar{\tc})$ results are $r^E=0.257$, $\mu=2.10$ $\mathcal{Q}=-0.23$.
 } 
\vspace{0 cm}
\begin{tabular}{|c@{\extracolsep{-0.6 cm}}c@{\extracolsep{0.2 cm}}ccccc|}
\hline
\hline
\rule{0ex}{3.0ex}
& & $r^{E}$ & $r^{M}$ & $r^{Q}$ & $\mu$ & $\mathcal{Q}$  \\ 
\hline
 \rule{0ex}{3.0ex}
 \multirow{2}{2.0cm}{$B_{\tc}^{0*}(\tc\bar{\tb})$ } & Our  & 0.290 & 0.564 & 0.382 & 2.95 & $-1.37$
\\ 
\rule{0ex}{3.0ex}
& Ref. \cite{Simonis:2016pnh}  & -& - & - & 0.35 & - \\
\rule{0ex}{3.0ex}
 & Ref. \cite{Xu:2024fun}  &0.23 &- & - &3.19 & $-$0.67     \\
\hline
 \rule{0ex}{3.0ex}
 \multirow{1}{2.0cm}{$\Jpsi(\tc\bar{\tc})$ } & Our & 0.350 & 0.492 & 0.275 & 2.03 & $-0.70$
\\ 
\rule{0ex}{3.0ex}
& Ref. \cite{Bhagwat:2006pu}  &  0.052 &- &- & 2.13 & $-0.28$ \\
\rule{0ex}{3.0ex}
& Ref. \cite{Xu:2019ilh} & 0.24 &- &- & 2.12  & $-0.33$ \\
\rule{0ex}{3.0ex}
& Ref. \cite{Dudek:2006ej} &0.066 &- &- & 2.10  & $-0.23$ \\
\rule{0ex}{3.0ex}
& Ref. \cite{Adhikari:2018umb} &0.045 &- &- & 1.952  & $-0.78$ \\
\hline
 \rule{0ex}{3.0ex}
 \multirow{1}{2.0cm}{$\Upsilon(\tb\bar{\tb})$ } & Our & 0.279 & 0.395 & 0.217 & 2.01 & $-0.69$
\\ 
\rule{0ex}{3.0ex}
& Ref. \cite{Adhikari:2018umb} &0.016&- &- & 1.985  & $-0.731$ \\
\hline
\hline
\end{tabular}
\end{center}
\end{table}

Nonetheless, wherever possible, we compare with the available data. We highlight that for a structureless spin-$1$ particle where the magnetic and quadrupole moments take the value $ \mu = 2$ and
$\mathcal{Q}=-1$ 
\cite{Brodsky:1992px}, any deviation from these values points to the internal structure of mesons.
From the tabulated results, we can infer that the heavy mesons are nearly point-like within the CI, as expected and as was reported in Ref.~\cite{Raya:2017ggu}. However, the heavy-light system deviates from this premise.
The charge radii of V mesons are consistently larger than PS mesons.
To analyze the behavior of the electric FFs of V mesons at large $Q^2$, we use a  parametrization similar to the EFFs of S and PS mesons, i.e., 
\begin{align}
G_{E}^{ V}(Q^2) = \frac{e_M+a_V^E Q^2 + b_V^E Q^4}{1+ c_V^E Q^2 + d_V^E Q^4} \, .
\label{eqE}\end{align}
In analogy, in the case of magnetic and quadrupole FFs, the parametrizations adopted are 
\begin{eqnarray} \label{eqMQ}\
G_{M}^{V}(Q^2) &=& \frac{\mu_V + a_V^M Q^2 +b_V^M Q^4}{1+ c_V^M Q^2 + d_V^M Q^4} \, , \\
G_{Q}^{V}(Q^2) &=& \frac{\mathcal{Q}_{V} +a_V^Q Q^2 + b_V^Q Q^4}{1+ c_V^Q Q^2 +d_V^Q Q^4} \, ,
\end{eqnarray}
where $a_i^V$, $b_i^V$, $c_i^V$ and $d_i^V$ with $i\in \{ E, M, Q\}$ are the coefficients to be fitted with our numerical results.
 In~Table \ref{tableVEMQ} we present these values for all the indicated coefficients of ground state V mesons. This parametric form is valid over the range $Q^2\in [0,8 M_M^2 ]$.
\begin{table*}[htbp]
\caption{\label{tableVEMQ}Parameters for the fits in Eqs.~(\ref{eqE})-(\ref{eqMQ}), for the electric, magnetic and quadrupole FFs of V mesons.}
 \renewcommand{\arraystretch}{1.6} %
\begin{tabular}{@{\extracolsep{0.1 cm}}|ccccccccccccc|}
\hline \hline
 & $a_V^E$ &  $b_V^E$ &  $c_V^E$ &  $d_V^E$ & $a_V^M$ &  $b_V^M$ &  $c_V^M$ &  $d_V^M$ & $a_V^Q$ &  $b_V^Q$ &  $c_V^Q$ &  $d_V^Q$   \\ 
\hline
$\rho(\tu\bar{\td})$ & $-0.113$& $-0.009$ &1.252 &0.013 &$-0.134$ & $-0.005$ & 1.085 &$-0.108$ & 0.032 &0.002&1.096&$-0.083$ \\
$K_1(\tu\bar{\ts})$ &   $-0.111$ & $-0.005$ &1.127&$-0.005$ & $-0.295$ &0.003 &0.774 &$-0.109$ &0.055 &0.002 &1.006& $-0.102$\\
$\phi(\ts\bar{\ts})$ &  $-0.215$ & 0.011 & 0.722 &$-0.083$ &$-0.171$ & $-0.003$ &0.749 & $-0.083$ & 0.075 & 0.001 & 0.714 &$-0.086$\\
$D^{*0}(\tc\bar{\tu})$ &  0.828 & 0.002 & 1.375 & 0.285 & 0.999 & 0.003 & 1.323 & 0.331 & 0.040 & 0.000 & 1.390 & 0.502\\
$D^*_{\ts}(\tc\bar{\ts})$ &  $-0.033$ & 0.000 & 0.506 & $-0.005$ & $-0.050$ & 0.000 & 0.678 & $-0.012$ & $-0.399$ & 0.001 & 1.413 & 0.450 \\
$\Jpsi(\tc\bar{\tc})$ &  0.972 & $-0.014$ & 1.425 & 0.715 & 0.990 & 0.005 & 0.977 & 0.378 & $-0.848$ & $-0.003$ & 1.643 & 0.711 \\
$B^{+*}(\tu\bar{\tb})$ & 0.145 & 0.000 & 0.483 & 0.010 & 0.705 & 0.003 & 0.894 & 0.043 & $-0.284$ & $-0.001$ & 0.919 & 0.055 \\
$B^{0*}_{\ts}(\ts\bar{\tb})$ &  0.145 & 0.000 & 0.373 & 0.029 & 0.453 & 0.000 & 1.204 & 0.085 & $-0.012$ & 0.000 & 1.396 & 0.041 \\
$B^{0*}_{\tc}(\tc\bar{\tb})$ & 0.067 & 0.000 & 0.298 & 0.006 & 0.431 & 0.000 & 0.542 & 0.022 & $-0.159$ & 0.000 & 0.507 & 0.020 \\
$\Upsilon(\tb\bar{\tb})$ & 0.047 & 0.000 & 0.387 & 0.006 & 0.140 & 0.000 & 0.443 & 0.009 & $-0.046$& 0.000 & 0.369 & 0.006 \\ 
\hline \hline
\end{tabular}
\end{table*}
Regarding the fitting parameters of electric FFs of V mesons, we notice that it shows the same tendency as the S and the PS mesons; $b_V^E$ is smaller than the other {\em electric} parameters; thus, the $1/Q^2$ behavior dominates for large $Q^2$. However, we point out that the $a_V^E$ coefficients are sizeable and negative for the mesons entirely composed by a combination of two of the three light quarks. Hence, these coefficients are responsible for the electric FFs of these mesons becoming negative faster as a function of $Q^2$ than the electric FFs of other mesons. Similarly, for the magnetic and quadrupole FFs of V mesons, we find that $b_V^M$ and $b_V^Q$ are relatively small. Therefore, the $1/Q^2$ behavior also prevails in these cases for the CI.

\begin{figure*}[t!]
\begin{tabular}{@{\extracolsep{-2.3 cm}}cc}
\hspace{-2cm}
\includegraphics[scale=0.65]{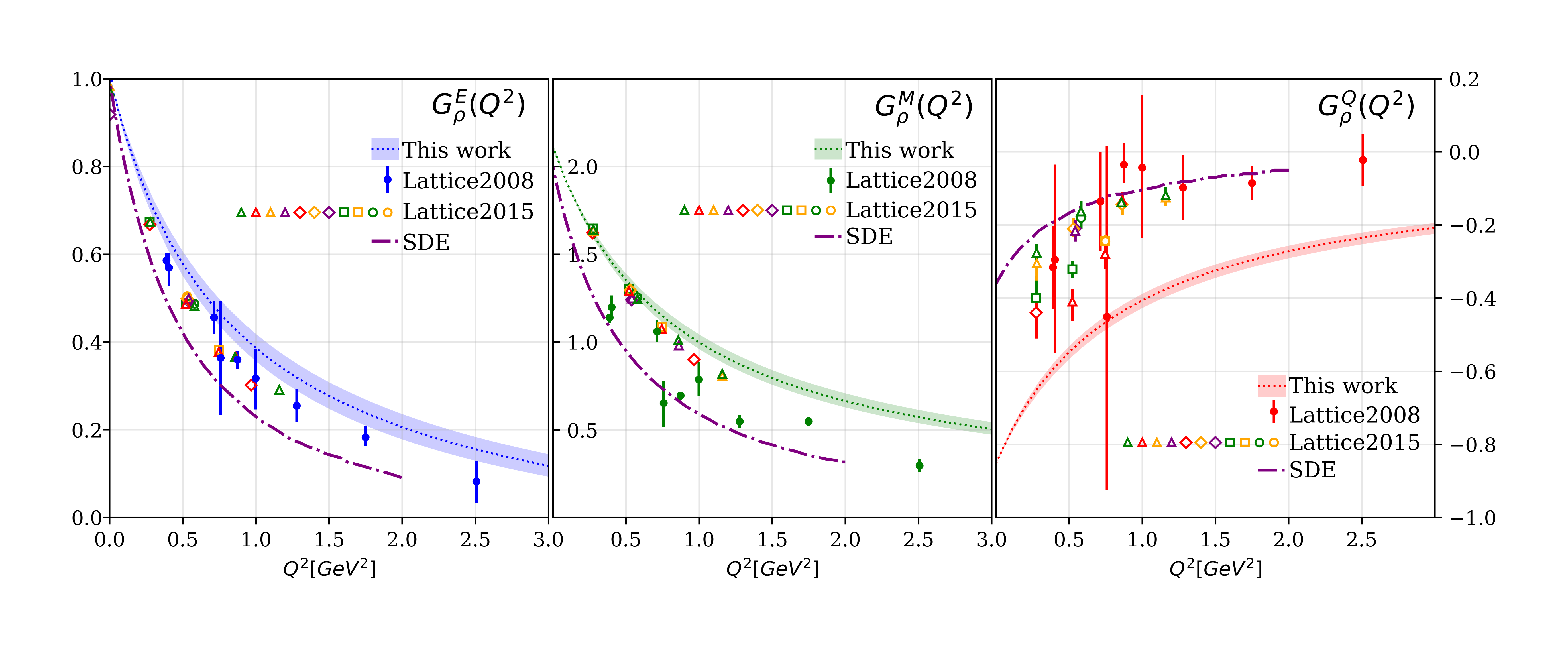}
\vspace{-1cm}
\end{tabular} 
    \caption{Electromagnetic (blue), Magnetic (green), and quadrupole (red) FFs for $\rho$-meson. The central curve in each case is obtained using the $\tau_{\rm UV}$ value from the \tab{parameters1}. The width of the band represents a $5\%$ variation in the charge radius. The meson life is so short that it is challenging to carry out experimental measurements of its EFFs. However, we compare our results with the ones obtained from lattice QCD~\cite{QCDSF:2008tjq,Shultz:2015pfa} and the 
    SDE~\cite{Xu:2024fun}.} \label{rho}
\end{figure*}

In Fig.~\ref{rho}, we plot the electric, magnetic and quadrupole FFs for the $\rho$-meson, allowing for a $5\%$ variation in the value of the charge radius of the electric FFs. It is primarily controlled by varying $\tau_{\rm UV}$.
The behavior of the $\rho$-meson EFFs is, as expected, in perfect agreement with the previous work reported with the CI~\cite{Roberts:2011wy}. We also plot the magnetic and quadrupole FFs. We emphasize that all curves exhibit a pole at $Q^2 = -M_V^2$ because of the behavior of the quark-photon vertex in the time-like region. We also display an explicit comparison with the earlier results reported by lattice groups~\cite{QCDSF:2008tjq,Shultz:2015pfa}. As we anticipated and expected, the CI results are harder due to its point-like interaction. This behaviour is typical of the CI model and it is irrespective of the nature of the mesons under study. In the next section, we continue with the discussion of the numerical results in greater detail for all V mesons, from the light to the heavy sector.
 \begin{figure}
    \centering
    \includegraphics[scale=0.44]{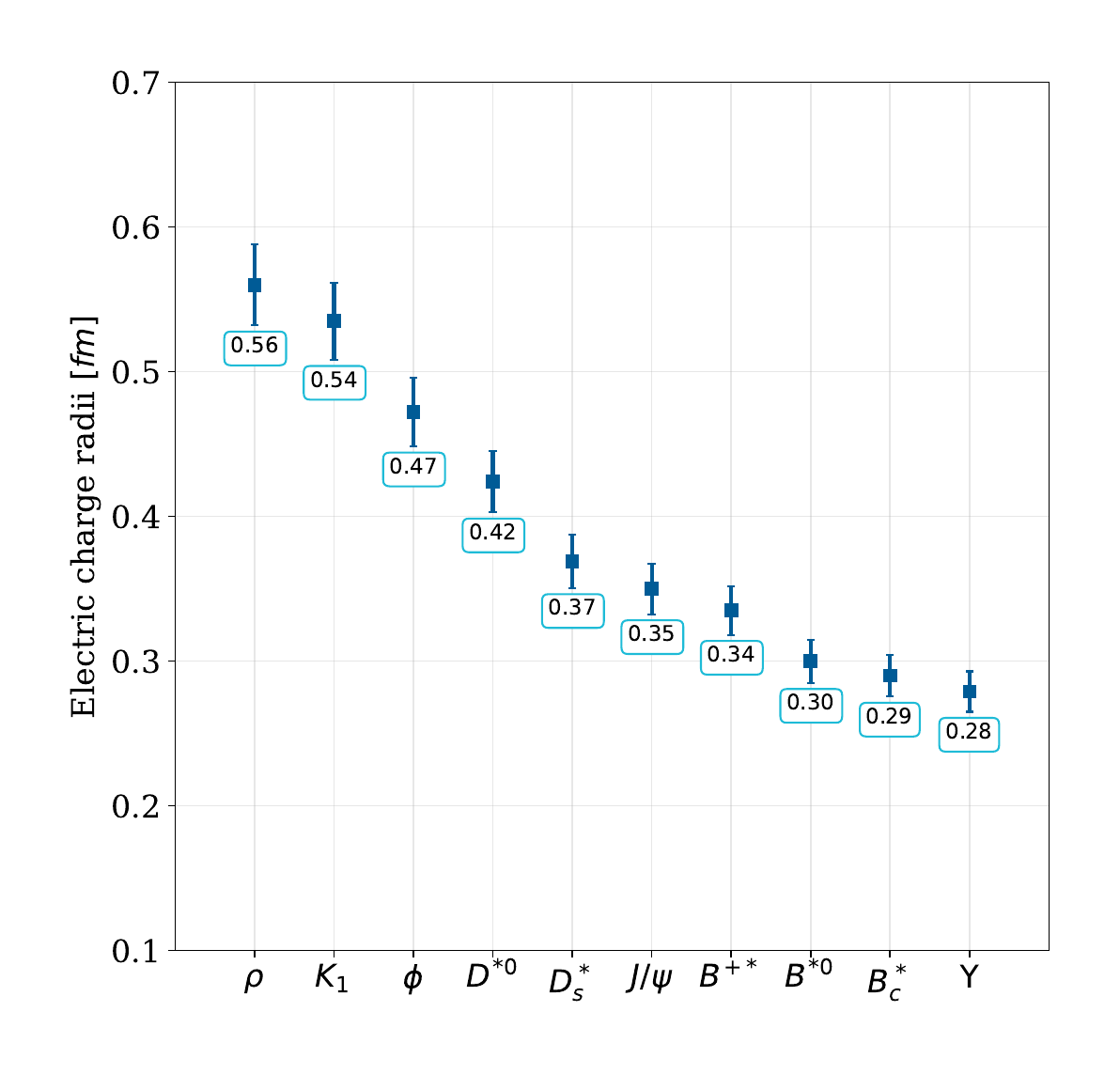}
    \vspace{-1cm}
    \caption{Charge radii of ground state V mesons in the CI computed using Eq.~(\ref{fradii}). It is clear from this plot that the radii tend to decrease as the constituent quarks' masses increase. This significant quality will be addressed in Sec. \ref{dis}.
    }
    \label{jerarquia1}
\end{figure}



\section{Discussion of Numerical Results}
\label{dis}
 After outlining the key features of our CI model, choice of appropriate parameters in agreement with our previous works and carrying out explicit numerical computation, we summarize our findings and make explicit comparisons with similar works carried out in the literature before\,: 
\begin{itemize}
\item  We plot electric, magnetic and quadrupole FFs of the V mesons $\rho$ $(\tu\bar{\td})$,
$K_1$ $(\tu\bar{\ts})$, 
$\phi$ $(\ts\bar{\ts})$, $\Jpsi(\tc\bar{\tc})$, $D^{*0}(\tc\bar{\tu})$,  $D_{\ts}^{*}(\tc\bar{\ts})$, $B_{\tc}^{*}(\tc\bar{\tb})$, $\Upsilon(\tb\bar{\tb})$, $B^{+*}(\tu\bar{\tb})$,  $B_{\ts}^{0*}(\ts\bar{\tb})$ in Fig. \ref{plotVE}. The electric FF has a zero before $Q^2=10$ GeV$^2$ for the lightest   $\tu\bar{\td}$, $\tu\bar{\ts}$ and $\ts\bar{\ts}$ mesons.
The zeros in the $G^E(Q^2)$ are located as follows\,:
\bea
\begin{tabular}{rccc}
$\rho(\tu\bar{\td})$: & & $G^E(Q^2)=0$  &  $\text{for}\;\; x=6.35$\, , \\
$K_1(\tu\bar{\ts})$: & & $G^E(Q^2)=0$   &  $\text{for}\;\; x=6.65$\, ,\\
$\phi(\ts\bar{\ts})$: & & $G^E(Q^2)=0$  &  $\text{for}\;\; x=7.59$\, ,
\end{tabular}
\eea
 where $x=Q^2/M_M^2$. Several previous studies suggest the existence of a zero in $G^E_{\rho}$ for $x\sim 6$~\cite{Brodsky:1992px, Haberzettl:2019qpa,Roberts:2011wy, Xu:2019ilh}.
 In Ref.~\cite{Roberts:2011wy}, it is shown that $G^E_{\rho}(Q^2)$ exhibits a zero at $Q^2 = 5.0$ GeV$^2$ within a CI. In the case of the electric FF of the positively charged $K_1$, Ref.~\cite{Xu:2019ilh} predicts a zero in $G^E_{K_1}(Q^2)$ at $x=8$, However, for the FF of the neutral $K_1$,  $G^E_{K_1}(Q^2)$ is positive definite on $x > 0$.
 The value of $x$ where $G^E(Q^2)$ crosses the zero is higher as the mass of the meson increases. It is because the $Q^2$ dominance sets in at its higher values.
Experimental data from JLab suggest that the proton’s electric FF might pass through zero at $x\sim 10$~\cite{Jones:1999rz, Gayou:2001qd, Punjabi:2005wq, Puckett:2010ac, Puckett:2011xg}. Focusing our attention on the V meson's electric FFs,  $G_V^E(Q^2)$ can serve as a surrogate for the proton’s electric FF. Since the dynamical reasons for the potential appearance of a zero is not expected to be dissimilar in both cases, analyses of $G_V^E(Q^2)$ may provide us with qualitatively sound guidance on the possible appearance and location of a zero in proton’s EFF.
\item 
Regarding the $\rho$ meson, FFs, and their corresponding charge radii as well as magnetic and quadrupole moments agree with the first results employing the CI~\cite{Roberts:2011wy}. As in the case for S and PS mesons, we allow for a $5\%$ variation of the $\rho$ charge radius to chart out its effect on the subsequent evolution of the $G_\rho^E$,$G_\rho^M$ and $G_\rho^Q$ as a function of $Q^2$, Fig.~\ref{rho}. A small variation of the initial slope of the curve $Q^2 \simeq 0$ opens a noticeable spread for large $Q^2$ but keeps the qualitative and quantitative behavior largely intact. The results for the static properties of the $\rho$ meson are as follows:
charge radius $r^E= 0.560$ fm, magnetic moment $\mu = 2.11$, and quadrupole moment $\mathcal{Q}=-0.85$.
For $\rho(\tu\bar{\td})$, our results show a deviation of $5.21\%$ for magnetic and $15\%$ for quadrupole moments in comparison with the expected values of {\em structureless} V mesons. 
\item
With our CI model, we show the obtained ratios for $\mu/\mathcal{Q}$ in Table~\ref{muyq}.
\begin{table}[b!]
    \centering
   \begin{tabular}{@{\extracolsep{0.2cm}}|ccc|ccc|}\hline\hline 
\,\, Meson &  $\mu/\mathcal{Q}$ & \% & Meson & $\mu/\mathcal{Q}$ &\% \\
\hline
$\rho(\tu\bar{\td})$ &  $-$2.48 & 19.35 \, & $B^{+*}(\tu\bar{\tb})$ &$-$1.84 & 8.70\, \\
$K_1(\tu\bar{\ts})$ &  $-$2.42 & 17.36 \,& $B^{0*}_{\ts}(\ts\bar{\tb})$ &$-$1.51 & 32.45\, \\
$\phi (\ts\bar{\ts})$ &  $-$2.51 & 20.32 \, & $B^{0*}_{\tc}(\tc\bar{\tb})$ &$-$2.15 & 6.98\,\\
$B^{+*}(\tc\bar{\tu})$ &  $-$1.43 & 39.86\, & $\Jpsi(\tc\bar{\tc})$ &$-$2.90 & 34.98\,\\
$B^{0*}_{\tc}(\tc\bar{\ts}$) &  $-$2.95 & 32.20\, & $\Upsilon(\tb\bar{\tb})$ &$-$2.90 & 34.98\,\\\hline
\hline
\end{tabular}
    \caption{\label{muyq}The ratio between the magnetic and quadrupole moment for the V mesons is discussed in this work. The final column displays the percentage deviation from the value predicted for {\em structureless} V mesons.}
    \label{tab:my_label}
\end{table}
It is noticeable that the lighter mesons are closer to the value of  $\mu/\mathcal{Q}=-2$ expected of {\em structureless} V mesons. The maximum percentages of difference with this value are for the heavy and heavy-light mesons.
\item In Tables~\ref{fbo},~\ref{fbo1} and~\ref{fbo2}, we show our predictions for the  electric, magnetic, and quadrupole radii. We also compare the results obtained with this formalism with other available studies. In constituent like quark models, $\mathcal{Q}_\rho = G_\mathcal{Q}(Q^2 = 0) < 0$ corresponds to oblate deformation \cite{Krivoruchenko:1985vz}. This behavior for the quadrupole moment is evident in all mesons except in the cases of the charged V mesons, namely, $B^{0*}_{\ts}(\ts\bar{\tb})$ and $D^{*0}(\tc\bar{\tu})$.
\item
It is interesting to see the overall trend of decreasing charge radii with increasing constituent quark mass, as shown by the hierarchy below,
\begin{eqnarray}
&& r_{\tu\bar{\td}} > r_{\tu\bar{\ts}} > r_{\tc\bar{\tu}} > r_{\tu\bar{\tb}} \,, \nn \\
&& r_{\tu\bar{\ts}} > r_{\ts\bar{\ts}} > r_{\tc\bar{\ts}} > r_{\ts\bar{\tb}} \,, \nn \\
&& r_{\tc\bar{\tu}} > r_{\tc\bar{\ts}} > r_{\tc\bar{\tc}} > r_{\tc\bar{\tb}} \,, \nn \\
&& r_{\tu\bar{\tu}} > r_{\ts\bar{\ts}} > r_{\tc\bar{\tc}} > r_{\tb\bar{\tb}} \,.\nn 
\end{eqnarray}
This behavior is shown pictorially in Fig.~\ref{jerarquia1}. The magnetic moment is almost independent of the quark mass.
We must emphasize and reiterate that the CI model is only a simple model to capture the infrared features of strong interactions. Refined QCD calculations are required to shed further light on these findings and reaffirm or improve upon the results presented in this work.
\item 
The charge radii of the V mesons are larger than the PS mesons according to the calculations carried out with this model. A comparison of the charge radii of V mesons with their PS analogues is presented in Table \ref{radiipsyv}. 
The maximum difference is $74\%$ and occurs for the heaviest meson ($\Upsilon(\tb\bar{\tb})$), while the minimum is $1.5\%$ for $B^{+*}(\tu\bar{\tb})$.

\begin{table}[htb]
\caption{\label{radiipsyv} Charge radii for PS and V mesons. The results of the charge radii of the PS mesons are taken from a previous work using the same model~\cite{Hernandez-Pinto:2023yin}. In the last column we show the percentage of difference between these states.}
 \renewcommand{\arraystretch}{1.6} %
\begin{tabular}{|@{\extracolsep{0.3 cm}}cccc|}
\hline \hline
Meson & $r_V^E$ & $r_{PS}^E$ \cite{Hernandez-Pinto:2023yin} & difference (\%)  \\ 
\hline
$\rho(\tu\bar{\td})$ & 0.560 & 0.45 & 19 \\
$K_1(\tu\bar{\ts})$ & 0.535 & 0.42 & 21  \\
$\phi(\ts\bar{\ts})$ & 0.472 & 0.36 & 23 \\
$D^{*0}(\tc\bar{\tu})$ &  0.424 & 0.36 & 15\\
$D^{*}_{\ts}(\tc\bar{\ts})$ & 0.369 & 0.26 & 29  \\
$B^{+*}(\tu\bar{\tb})$ & 0.335 & 0.34 & 1.5\\
$B^{0*}_{\ts}(\ts\bar{\tb})$ & 0.300 & 0.24 & 20 \\
$B^{0*}_{\tc}(\tc\bar{\tb})$ & 0.290  & 0.17 & 41  \\
$\psi(\tc\bar{\tc})$ &  0.350 & 0.20 & 42\\
$\Upsilon(\tb\bar{\tb})$ & 0.279 & 0.07 & 74  \\ 
\hline \hline
\end{tabular}
\end{table}
\end{itemize}
This brings us to conclude our in-depth examination of FFs using 
 CI for the ground states V mesons composed up of light and heavy mesons.
 \section{Conclusions}
\label{Conclusions}
In conclusion, we have carried out the following tasks in the study of ground state V mesons in the light and the heavy sector\,:

\begin{itemize}

\item

We compute the EFFs $G^E(Q^2),G^M(Q^2),G^Q(Q^2)$ of ten vector mesons composed of both light and heavy quarks in a CI model, Fig.~\ref{plotVE}. 

\item 

Furthermore, we analyze the sensitivity of $G^E_{\rho}(Q^2)$, $G^M_{\rho}(Q^2)$ and $G^Q_{\rho}(Q^2)$ in their $Q^2$ evolution by a slight permissible change in appropriate parameters to allow for a 5\% variation in the electric charge radius.

\item 

Interpolation functions are provided in 
Eqs.~(\ref{eqE})-(\ref{eqMQ}) and  the fitted parameters are reported in
Table~\ref{tableVEMQ}; this exercise allows for a convenient algebraic analysis of the behavior of the EFFs in the momentum range that we mentioned above, for understanding the anticipated power law behavior in the asymptotic limit and for any application the reader may deem useful in future.

\item 

In continuation of our previous work for S and PS mesons~\cite{Hernandez-Pinto:2023yin} and as an ongoing contribution to a broader effort summarized and promised in~\cite{Aguilar:2024otb}, this study presents a unified treatment in calculating the EFFs of any meson in general, a formalism which can easily 
be extended to study AV mesons.

\item

As we expected, charge radii of the V mesons are larger than the charge radii of the PS mesons, Table~\ref{radiipsyv}. The quadrupole FF of the $\rho$ meson
is negative which implies that the distribution of charge in the V mesons is oblate. It was already known about the $\rho$-meson. This revealing feature is repeated for several other mesons that we study in this work.

\item 

The radii associated with the magnetic and quadrupole FFs have in general not been reported using other models. We present them in our work for the sake of completeness in 
Tables~\ref{fbo},~\ref{fbo1} and~\ref{fbo2}.
\item 

We expect the new reported EFFs with the CI in this article to be harder than the exact QCD predictions. However, this computation
should serve as a guide and as a useful first step towards the construction of more realistic models of the strong interaction to compute all related physical observables.

\end{itemize}

Although our formalism is ready and complete to investigate axial vector mesons, notice that this study is still challenging. Since there are hardly any experimental results and unlike vector mesons, there are not enough theoretical efforts that can adequately guide us to set our parameters phenomenologically. However, this work is currently under way. Finally, in the near future we plan to extend this study to diquarks, which in turn represents an initial step to the eventual study of baryon EFFs using this formalism. This study will provide numerous guiding outlines for constraining the parameters of the model and to directly contrast with existing and planned experiments for the baryons.

\begin{acknowledgements}
L.~X.~Guti\'errez-Guerrero gratefully acknowledges the National Council of Humanities, Sciences, and Technologies (CONAHCyT) for the support provided to her through the {\em C\'atedras} CONAHCyT program and Project CBF2023-2024-268, Hadronic Physics at JLab: Deciphering the Internal Structure of Mesons and Baryons, from the 2023-2024 frontier science call. The work of R.~J.~Hern\'andez-Pinto is supported by CONAHCyT (Mexico) Project No. 320856 ({\em Paradigmas y Controversias de la Ciencia 2022}), {\em Ciencia de Frontera 2021-2042} and {\em Sistema Nacional de Investigadores}.
 A.~Bashir wishes to acknowledge the {\em Coordinaci\'on de la Investigaci\'on Cient\'ifica} of the{\em Universidad Michoacana de San Nicol\'as de Hidalgo}, Morelia, Mexico,  grant no. 4.10, the {\em Consejo Nacional de Humanidades, Ciencias y Tecnolog\'ias}, Mexico, project CBF2023-2024-3544
as well as the Beatriz-Galindo support during his current scientific stay at the University of Huelva, Huelva, Spain. M.~A.~Bedolla acknowledges support from the Munich Institute for Astro-, Particle and BioPhysics (MIAPbP), funded by the {\em Deutsche Forschungsgemeinschaft} (DFG, German Research Foundation) under Germany's Excellence Strategy – EXC-2094 – 390783311.

\end{acknowledgements}
\appendix
\setcounter{equation}{0}
\setcounter{figure}{0}
\setcounter{table}{0}
\renewcommand{\theequation}{\Alph{section}.\arabic{equation}}
\renewcommand{\thetable}{\Alph{section}.\arabic{table}}
\renewcommand{\thefigure}{\Alph{section}.\arabic{figure}}

\section{Form Factor Formulae}
\label{App:EM}
We present the analytic expression for all coefficients needed in this appendix in order to determine the EFFs of V meson in the CI model.
For V mesons, the $\mathcal{A}_i^{\Me}$ functions takes the form,
\bea
\nn\mathcal{A}^{\Me}_1 &=& 2-\alpha \, ,\\
\nn \mathcal{A}^{\Me}_2 & =& \frac{1}{2} \left( \frac{\alpha(2\beta-1)Q^2}{M_M^2}+\alpha(10\beta - 7)-4\right) \, , \\
\nn \mathcal{A}^{\Me}_3 &=& \frac{2\alpha(1-2\beta)(Q^2 + 5M^2_{M} )}{Q^2+4 M^2_{M}} \, , 
\eea
while $\mathcal{B}_i^{\Me}$ are,
\bea
\nn \mathcal{B}^{\Me}_1 &=& 2((\alpha-2)\, \alpha^2 \, (1-\beta) \, \beta \, Q^2 +\alpha \, M_{\fd}^2 \nn \\
\nn&+&
2\, (1-\alpha)\,M_{\fd} \, M_{\fu} +(1-\alpha)^2 \, \alpha \, M^2_{M}) \, ,\\
\nn \mathcal{B}^{\Me}_2  &=& \frac{\alpha  Q^2 \left((2 \beta -1)M_{\fd} M_{\fu}\right)} {M_{M} ^2}\\
\nn &+&\frac{\alpha^2 Q^2 
   M_{M} ^2 \left(2 \alpha  \beta ^3-5 \alpha  \beta ^2+(\alpha +2)
   \beta +\alpha -1\right)}{M_{M} ^2}\nn \\
   \nn &+&\alpha  (6 \beta -5)
   M_{\fd} ^2+2 M_{\fd} M_{\fu}  (2 \alpha  \beta -\alpha -2) \nn \\
   \nn&-&(\alpha
   -1) \alpha  M_{M} ^2 (10 \alpha  \beta -7 \alpha -6 \beta +1) \, , \\
\nn \mathcal{B}^{\Me}_3  &=&\bigg[ 4 \alpha  M_{M} ^2 \left((3-6 \beta ) M_{\fd} ^2+2 (1-2  \beta ) M_{\fd} M_{\fu} \right. \\
\nn &+&(\alpha -1) M_M^2 \left(16 \alpha \beta ^2-6 (\alpha +1) \beta -5 \alpha +3 \right)) \\
\nn &-& 4 \alpha Q^2 \bigg((2 \beta -1) \left(M_{\fd} M_{\fu} +\alpha \beta  M_M^2 (\alpha  (\beta -3)+2)\right) \\
\nn  &+&(\alpha -1) \alpha M_{M} ^2\bigg)\bigg]\frac{1}{(Q^2+4M_{M}^2)}\, .
\eea

\bibliography{ccc-a}
\end{document}